\documentclass[10pt,conference]{IEEEtran}
   \usepackage{epsfig}
   \usepackage{color}
   \usepackage{latexsym}
   \usepackage{pslatex}
  \usepackage{algorithm}
   \usepackage{algorithmic}
   \usepackage{multirow}

\begin{document}

\title{Embedding of Deterministic Test Data for In-Field Testing}
\author{Nan Li ~~~~~ Elena Dubrova \\
Royal Institute of Technology (KTH), Forum 120, 164 40 Kista, Sweden\\
\{nan3,dubrova\}@kth.se
}

\maketitle

\begin{abstract}
This paper presents a new feedback shift register-based method for embedding deterministic test patterns
on-chip suitable for complementing conventional BIST techniques for in-field 
testing. Our experimental results on 8 real designs show that the presented 
approach outperforms the bit-flipping approach by 24.7\% on average.
We also show that it is possible to exploit the uneven distribution of 
don't care bits in test patterns in order to reduce the area required for 
storing deterministic test patterns more than 3 times with less than 2\% 
fault coverage drop.
\end{abstract}

\begin{keywords}
BIST, top-off test patterns, feedback shift register, NLFSR, in-filed testing.
\end{keywords}

\section{Introduction}

Large test data volume is widely recognized 
as a major contributor to the testing cost of integrated circuits~\cite{WaC05}.
The test data volume in 2017 is expected to be 10 times larger
than the one in 2012~\cite{ITRS}. On the contrary, the size of the Automatic Test 
Equipment (ATE) memory is expected to grow only twice~\cite{ITRS}.

A number of efficient on-chip test compression techniques have been proposed
as a solution for reducing ATE memory requirements, including~\cite{WaC05,KoB01,RaTKM04,MiK04,CzM11}. 
A test set for the circuit under test is 
compressed to a smaller set, which is stored in ATE memory. An on-chip decoder is used to generate the original test set from the compressed one during test application.
Test compression has already established itself as a mainstream design-for-test
methodology for manufacturing testing~\cite{CzM11}. 
However, it cannot be used for in-field testing where ATE is not available~\cite{MaA10}.

For in-field testing, Built-In Self Test (BIST) including use of JTAG 
is applied, in which either pseudo-random test patterns are generated 
within the system or pre-computed deterministic test patterns are stored 
in system memory~\cite{MaA10}.  
In terms of test application
time and fault coverage, deterministic test patterns are obviously more effective 
than pseudo-random ones. The fault coverage achieved with  pseudo-random test patterns
can be as low as 65\%~\cite{DaT00}. 
Several methods for increasing BIST test coverage have been proposed, including modification of the circuit under test~\cite{EiL83}, insertion of control and observe points into the circuit~\cite{RaTKM04}, modification of the LFSR to generate a
sequence with a different distribution of 0s and 1s~\cite{ChM84},
embedding of deterministic test patterns into LFSR's patterns by 
LFSR re-seeding~\cite{Jer08} or bit-flipping~\cite{WuK96}, or storing 
them in an on-chip memory~\cite{SaDB84}. The idea of complementing pseudo-random patterns with deterministic patterns 
is particularly attractive because the deterministic patterns can also solve 
the problem with transition or delay faults which are not handled efficiently by the
pseudo-random patterns. 
However, the area required to store deterministic test patterns
within the system can be prohibitively high.
For example, the memory required to store them may exceed 30\% of the memory used in a conventional ATPG approach~\cite{HeF99}.


In this paper, we propose a new method for embedding deterministic test patterns
on-chip suitable for complementing conventional techniques for in-field testing. 
We generate deterministic test patters using a structure known as {\em binary machine}. This name was introduced by S. Golomb in his seminal book~\cite{Golomb_book}.
Binary machines can be considered as a more general type of 
Non-Linear Feedback Shift Registers (NLFSRs)~\cite{Ja89}
in which every stage is updated by its own feedback function. 

Binary machines are typically smaller and
faster than NLFSRs generating the same sequence. 
For example, consider the 4-stage NLFSR with the feedback function 
\[
f(x_0,x_1,x_2,x_3) = x_0 \oplus x_3 \oplus x_1 \cdot x_2 \oplus x_2 \cdot x_3,
\]
where "$\oplus$" is the XOR (addition modulo 2), "$\cdot$" is the AND,
and $x_i$ is the variable representing the value of the stage $i$, 
$i \in \{0,1,2,3\}$.
If this NLFSR is initialized to the state $(x_0 x_1 x_2 x_3) = (1000)$, it 
generates the output sequence 
\begin{equation} \label{bs}
(1,0,0,0,1,1,0,1,0,1,1,1,1,0,0)
\end{equation}
with the period 15.
The same sequence can be generated by the 4-stage binary machine with the 
feedback functions
\[
\begin{array}{lcl}
f_3(x_0,x_3) & = & x_0 \oplus x_3 \\
f_2(x_1,x_2,x_3) & = & x_3 \oplus x_1 \cdot x_2  \\
f_1(x_2) & = & x_2 \\
f_0(x_1) & = & x_1.
\end{array}
\]
We can see that the binary machine uses 3 binary operations,
while the NLFSR uses 5 binary operations. Furthermore,
the depth of feedback functions of the binary machine is smaller than
the depth of the feedback function of the NLFSR. Thus, the
binary machine has a smaller propagation delay than the NLFSR.

While binary machines can potentially be smaller and faster than NLFSRs, 
the search space for finding a best binary machine for a given sequence
is much larger than the corresponding one for NLFSRs.
Algorithms for constructing binary machines 
were presented in~\cite{Du10aj,Du11a}. Both algorithms
result in binary machines with the minimum number of stages
for a given binary sequence. However, they do not minimize the circuit 
complexity of feedback functions.
For Finite State Machines (FSM), it is known that an FSM 
with a non-minimal number of stages, e.g. encoded
using one-hot encoding, often has a smaller total size
than an FSM with a minimal number of stages~\cite{MiBS85}.

In this paper, we present an algorithm with constructs binary machines
with a non-minimal number of stages.
Our experimental results show that binary machines 
constructed by the presented algorithm are 63.28\% smaller 
on average compared to the one constructed 
by the algorithm~\cite{Du11a}.
The presented algorithm is particularly efficient for incompletely specified sequences,
which are important for testing.

The rest of the paper is organized as follows. 
Section~\ref{bm} gives an introduction to binary machines.
Section~\ref{sa1}, describes the new algorithm for constructing binary
machines.  
Section~\ref{exp} presents the experimental results.
Section~\ref{con} concludes the paper and discusses open problems.

\section{Binary Machines} \label{bm}

\begin{figure}[t!]
\begin{center}
    \includegraphics*[clip=true,width=3.5in]{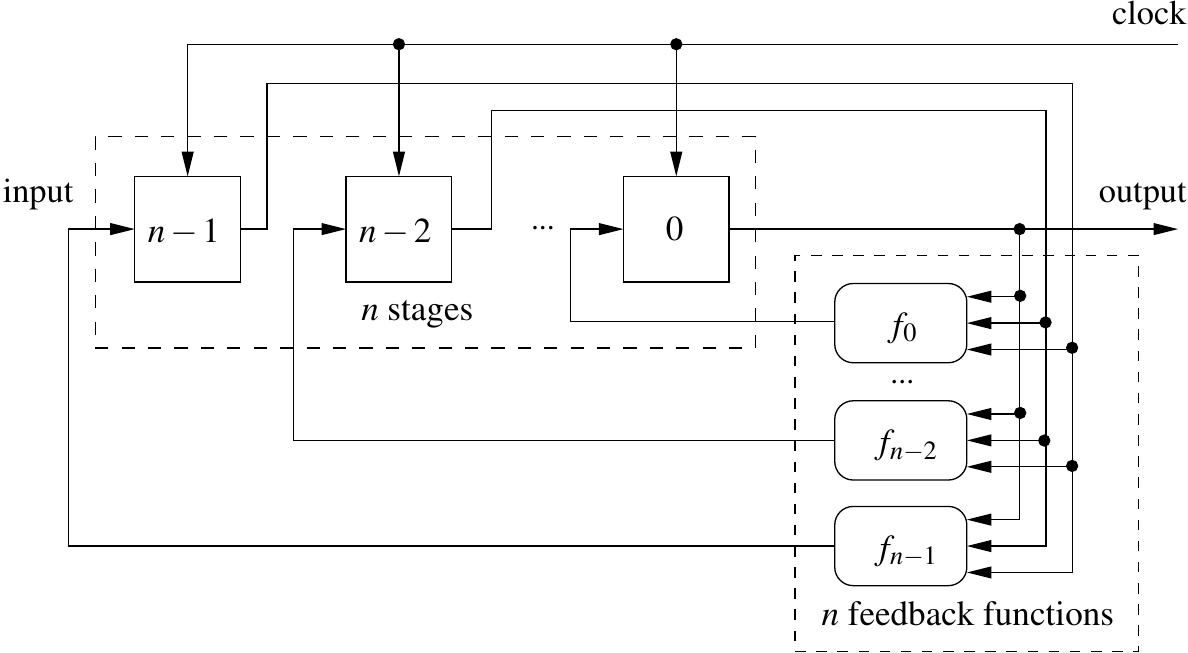}
\caption{The general structure of an $n$-stage binary machine.}\label{bin_machine}
\end{center}
\end{figure}

An $n$-stage binary machine consists of $n$
binary storage elements, called {\em stages}~\cite{Golomb_book}. 
Each stage $i \in \{0,1, \ldots,n-1\}$ has an associated {\em state variable} $x_i \in \{0,1\}$ 
which represents the current value of the stage $i$ and a {\em feedback function} 
$f_i: \{0,1\}^n \rightarrow \{0,1\}$ which determines how the value of $x_i$ is updated (see Figure~\ref{bin_machine}).

A {\em state} of a binary machine is a vector of values of its
state variables. At every clock cycle, 
the next state of a binary machine is determined from its current state 
by updating the values of all stages simultaneously
to the values of the corresponding feedback functions.
An $n$-stage binary machine has $2^n$ states corresponding to the 
set $\{0,1\}^n$ of all possible binary $n$-tuples.

The {\em degree of parallelization} of an $n$-stage binary machine, $k$, is the number of output bits  generated at each clock cycle, $1 \leq k \leq n$.



The {\em dependence set} of a Boolean function $f: \{0,1\}^n \rightarrow \{0,1\}$ is defined by
\[
dep(f) = \{j \ | \ f(X)|_{x_j=0} \not = f(X)|_{x_j=1}\},
\]
where $f(X)|_{x_j=k} = f(x_0, \ldots, x_{j-1}, k, x_{j+1}, \ldots, x_{n-1})$
for $k \in \{0,1\}$.


The {\em Algebraic Normal Form (ANF)}~\cite{CuS09} of a Boolean function $f: \{0,1\}^n \rightarrow \{0,1\}$ 
(also called {\em Reed-Muller canonical form}~\cite{Gr91})
is an expression in the Galois Field or order 2, $GF(2)$, of type
\[
f(x_0, x_1,\ldots,x_{n-1}) = \sum_{i=0}^{2^n-1}  c_i \cdot 
x_0^{i_0} \cdot x_1^{i_1} \cdot \ldots \cdot x_{n-1}^{i_{n-1}},
\]
where $c_i \in \{0,1\}$ are constants
and $(i_0 i_1 \ldots i_{n-1})$ is the binary
expansion of $i$.

\section{Related Work}~\label{prev}

The first algorithm for constructing a binary machine with the minimum number of stages for a
given binary sequence was presented in~\cite{Du10aj}. This algorithm 
exploits the unique property of binary machines that {\em  any} binary $n$-tuple can be the next state of 
a given current state. 
The algorithm assigns every 0 of a sequence a unique even integer and every 1 of a sequence a unique odd integer.
Integers are assigned in an increasing order starting from 0. For example, if an 8-bit sequence
00101101 is given, the sequence of integers 0,2,1,4,3,5,6,7 can be used. This sequence of integers is interpreted as
a sequence of states of a binary machine. The largest integer in the sequence of states
determines the number of stages. In the example above, $\lceil \log_2 7 \rceil = 3$,
thus the resulting binary machine has 3 stages. 
The feedback
functions $f_0, f_1, f_2$ implementing the resulting current-to-next state mapping
are derived using the traditional logic synthesis 
techniques~\cite{espr}. 

Note that, in general, any permutation of integers 
can be used as a sequence of binary machine's states, as long as the selected integer modulo 2 is equal to the corresponding bit of the output sequence.
Different state assignments result in different feedback functions. The size of these 
functions may vary substantially. 
 
In~\cite{Du11a}, the algorithm~\cite{Du10aj} was extended to binary machines generating $k$ bits 
of the output sequence per clock cycle. 
The main idea is to encode a binary sequence into an $m$-ary sequence which can be 
generated in a simpler way. As an example,
suppose that we use the 4-ary encoding (00) = 0, (01) = 1, (10) = 2, (11) = 3
to encode the binary sequence 00101101 from the example above into
the  quaternary sequence 0231.
Then, we can construct a parallel binary machine generating 00101101 2-bits per clock cycle with a sequence 
of states 0, 2, 3, 1. Note that $\lceil \log_2 3 \rceil = 2$, so the resulting parallel binary machine has one stage
less than the binary machine constructed above. 
This is surprising taking into account that all existing techniques for the 
parallelization of LFSRs~\cite{PeZ92,MuS06} and NLFSRs~\cite{canniere-trivium,hell-grain} have area penalty.
In was shown in~\cite{Du11a} that, for random sequences,  
parallel binary machines can be an order of magnitude smaller than parallel LFSRs or NLFSRs generating the same sequence.
 
\section{Synthesis of binary machines} \label{sa1}

The problem of finding a best binary machine for a given sequence can be divided into
three sub-problems:
\begin{enumerate}
\item Selecting an optimal degree of parallelization for a given binary sequence.
\item Choosing an optimal state assignment for a given degree of parallelization.
\item Finding a best circuit for feedback functions for a given state assignment. 
\end{enumerate}

\subsection{Optimal degree of parallelization}

The degree of parallelization determines how many output bits are
generated per clock cycle. 
The size of binary machines may differ substantially for different 
parallelization degrees.
The degree of parallelization is {\em optimal} if it minimizes the size 
of the resulting binary machine. 

In order to construct a binary machine with the degree of parallelization 
$p$, we map a binary sequence into an $2^p$-ary sequence by
partitioning the binary sequence into vectors of length $p$.
The resulting vectors are treated as binary expansions of elements 
of an $2^p$-ary sequence. The same approach was used in~\cite{Du11a}. 

Let us denote by $N_i$ the number of occurrences of a digit $i$ in the $2^p$-ary sequence,
$0 \leq i < 2^p$. Let $N_{max}$ be the largest of $N_i$. In~\cite{Du11a}, it was shown that the minimum number of stages in a binary machine generating a given binary sequence with the degree 
of parallelization $p$ is equal to 
\begin{equation} \label{min_st}
k = \lceil log_2 N_{max} \rceil + p.
\end{equation}

From~(\ref{min_st}) we can see that if $N_{max} = 1$,
then $k = p$. Such a case is called {\em full parallelization}. 
On the base of our experimental results, we hypothesise that
the optimal degree of parallelization belongs to the interval
\begin{equation} \label{opt_st}
 1 \leq p_{opt} \leq \lceil \log_2 n \rceil
\end{equation}
where $n$ is the sequence length.

Note that for some applications, including testing,
the degree of parallelization is specified by the user.
For example, for testing it is equal to the number of scan chains. 

\subsection{Optimal state assignment}

A state assignment determines a sequence of states which a binary machine follows.
Different sequences of states give raise to different current-to-next state
mappings and, thus, to different updating functions. 
The state assignment is {\em optimal} if it minimizes the size 
of the resulting binary machine. 

Since a binary machine is a deterministic finite state automaton, any current state has a unique next state.
For a given $2^p$-ary encoding, the minimal number of bits which 
has to be added to $p$-tuples  
to make the current-to-next state mapping unique
is $\lceil log_2 N_{max} \rceil$. The minimal number of stages 
in the resulting binary machine is given by~(\ref{min_st}).

The strategy for state assignment presented in this paper 
has two major differences from the one in~\cite{Du11a}.
First, we use a non-minimal number of stages, namely
\begin{equation}
k \geq \lceil log_2 \frac{n}{p} \rceil + p.
\end{equation}

Second, we assign states so that the feedback functions implementing the current-to-next
state mapping depend on the minimum number of state variables.
It is known that a Boolean function of $k$ variables 
needs $O(2^k/k)$ gates to be implemented (Shannon-Lupanov bound)~\cite{We87}.
Feedback functions of binary machines are random functions. 
For random functions, their actual size is very close to the upper bound.
So, each extra variable nearly doubles the size of the function.

In our method, the feedback functions of an $(m+p)$-stage binary
machine depend on $m = \lceil log_2 \frac{n}{p} \rceil$
variables only.
In~\cite{Du11a}, the feedback functions can potentially
depend on all state variables.

The pseudocode of the presented state assignment algorithm is shown as Algorithm~\ref{alg1}. The input of the algorithm in a binary sequence $A = (a_0, a_1, \ldots, a_n)$
and the desired degree of parallelization $p$.
The output is a sequence $S = (s_0, s_1, \ldots, s_{r-1})$ of binary vectors 
$s_i = (s_{i,0}, s_{i,1}, \ldots, s_{i,p+m-1}) \in \{0,1\}^{p+m}$, where $r = \lceil n/p \rceil$ and $m = \lceil log_2 r \rceil$, corresponding to the states of an $(p+m)$-stage binary machine generating $A$ with the degree of parallelization $p$.

The algorithm partitions $A$ into $p$-tuples and appends at the beginning of each $i$th
$p$-tuple $m$ extra bits. These extra bits correspond to the binary expansion 
of the $i$th element of the permutation vector $\Pi$. 

Next, we define a mapping $s_i \mapsto s_{i+1}$, for all $i \in
\{0,1,\ldots,r-2\}$. Since $\Pi$ is a permutation, each state 
in the resulting sequence of states has a unique next state,
so the mapping is well-defined. 
The last state $s_{r-1}$ and each of the $2^{p+m}-r$ 
remaining states of the resulting binary $(p+m)$-stage machine 
are mapped to don't cares values. This 
gives us the possibility to specify the functions $f_0, f_1, \ldots, f_{p+m}$
implementing the current-to-next state mapping in a way which 
minimizes their size.
Since $r \leq 2^m$, we can treat them as functions depending
on the first $m$ variables only.
This is very important, because, as we mentioned above, 
for random functions, the size 
nearly doubles with each extra variable.

Since, by construction, the first $p$ bits of each state $s_i$
in $S = (s_0, s_1, \ldots, s_{r-1})$ correspond to the $i$th
$p$-tuple of $A$, the resulting binary machine generates $A$ 
with the degree of parallelization $p$.

\begin{algorithm}[t!]
\caption{Assign states to a binary machine
which generates an binary sequence $A = (a_0, a_1, \ldots, a_n)$ 
with the degree of parallelization $p$.}
\label{alg1}
\begin{algorithmic}[1]
{\small
\STATE $r := \lceil n/p \rceil$
\STATE $m := \lceil log_2 r \rceil$
\STATE $\Pi := (\pi_0, \pi_1, \ldots, \pi_{2^m-1})$ is a permutation of $(0,1,\ldots,2^m-1)$
\STATE Let $\pi_{i,j}$ be the $j$th element of the binary expansion of $\pi_i$, $j \in \{0,\ldots,m-1\}$
\FOR{every $i$  from 0 to $r-1$}
\FOR{every $j$  from 0 to $p-1$}
\STATE $s_{i,j} := a_{i*p+j}$  
\ENDFOR
\FOR{every $k$  from 0 to $m-1$}
\STATE $s_{i,p+k} := \pi_{i,k}$  
\ENDFOR
\STATE $s_i := (s_{i,0}, s_{i,1}, \ldots, s_{i,p+m-1})$
\ENDFOR
\STATE Return $S = (s_0, s_1, \ldots, s_{r-1})$
}
\end{algorithmic}
\end{algorithm}

As an example, let us construct a binary machine 
which generates the following 20-bit binary sequence with the degree of parallelization 2:
\[
A = (0,0,1,1,0,1,1,1,0,0,1,0,1,1,1,0,1,1,0,0).
\]
Since $n = 20$ and $p = 2$, we get
$r = 10$ and $m = 4$. Suppose we use the following permutation
of $(0,1,\ldots,15)$:
\[
\Pi = (1, 8, 4, 2, 9, 12, 6, 11, 5, 10, 13, 14, 15, 7, 3, 0)
\]
Then, we get the following sequence of states:
\[
\begin{array}{l}
S = (000100, 100011, 010001, 001011, 100100, 110010, \\
~~~~~ 011011, 101110, 010111, 101000) 
\end{array}
\]
The functions implementing the resulting 
current-to-next state mapping have the following defining table:
\[
\begin{tabular}{|c|c@{}c@{}c@{}c@{}c@{}c|} \hline
$x_5 x_4 x_3 x_2$ & $f_5$ & $f_4$ & $f_3$ & $f_2$ & $f_1$ & $f_0$  \\ \hline
0~0~0~1 & 1&0&0&0&1&1 \\
1~0~0~0 & 0&1&0&0&0&1 \\
0~1~0~0 & 0&0&1&0&1&1 \\
0~0~1~0 & 1&0&0&1&0&0 \\
1~0~0~1 & 1&1&0&0&1&0 \\
1~1~0~0 & 0&1&1&0&1&1 \\
0~1~1~0 & 1&0&1&1&1&0 \\
1~0~1~1 & 0&1&0&1&1&1 \\
0~1~0~1 & 1&0&1&0&0&0 \\
1~0~1~0 & -&-&-&-&-&- \\ \hline
\end{tabular}
\]
where "-" stands for a don't care value.
Recall that the functions depend of the four variables $x_5, x_4, x_3, x_2$ only.
The remaining 6 input assignments are mapped to don't cares.
We can implement the above functions as:
\[
\begin{array}{l}
f_5 = x_2 \oplus x_3\\
f_4 = x_5\\
f_3 = x_4\\
f_2 = x_3\\
f_1 = x_2 \oplus x_4\\
f_0 = (x_2 \oplus x_3)' \oplus x'_3 x'_3 x'_3\\
\end{array}
\]
where "$'$" stands for a complement.

It is important to use permutations $\Pi$ which have a low-cost implementation.
Examples of such permutations are sequences of states generated by counters, LFSRs, or NLFSRs with simple feedback functions~\cite{Du12}. In the example above, we used the sequence of states of an LFSR with the generator polynomial $1+x+x^4$.

\subsection{Best circuit for feedback functions}

The problem of finding a best circuit for a given Boolean function is known 
to be notoriously hard.
The exact solutions are known only for up to five variable functions~\cite{Kn98}.
However, there are many powerful heuristic algorithms for multi-level circuit optimization which are capable of finding good circuits for larger functions~\cite{espr}. 

We optimize feedback functions using UC Berkeley's tool ABC~\cite{abc}.
Our experimental results show that, even for random functions, ABC is 
capable of reducing the size of the original, non-optimized circuit by 30\% on average.

\section{Experimental Results} \label{exp}

\subsection{Comparison to previous BM synthesis algorithms}

\begin{figure}[t!]
\begin{center}
    \includegraphics*[clip=true,width=3.5in]{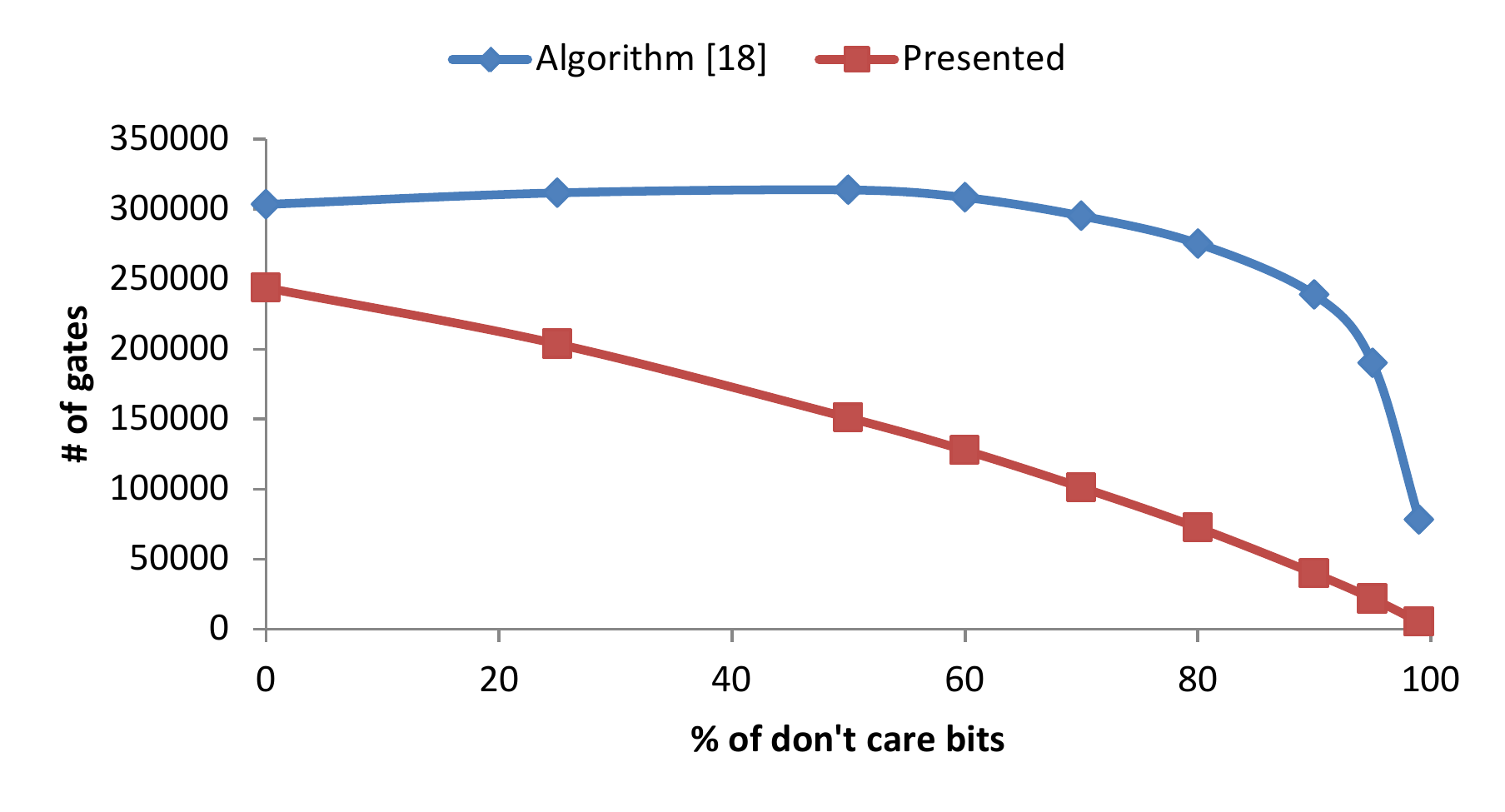}
\caption{Results for random sequences of length $2^{20}$.}\label{ff}
\end{center}
\end{figure}

In the first experiment, we compared the presented algorithm to the
algorithm~\cite{Du11a}. Using both algorithms, we constructed binary machines
for random sequences of length $2^{20}$ with a different number of don't care bits.
The results are summarized in Table~\ref{ta1} and Figure~\ref{ff}. 
As we can see, the presented
algorithm is significantly more efficient than the algorithm~\cite{Du11a}
for sequences with many don't cares. For the case of 99\% don't cares, it
outperforms the algorithm~\cite{Du11a} by 93.4\%.  


\begin{table}[t!]\centering\footnotesize
\begin{tabular}{|c|c|c|c|} \hline               
\% of  don't &  \multicolumn{2}{c|}{Number of gates in BMs} & \multirow{2}{*}{$\frac{G_1-G_2}{G_1}*100\%$} \\ \cline{2-3}
care bits & Alg.~\cite{Du11a}, $G_1$	& Presented, $G_2$ &    \\ \hline
0       &       303307  &       243734  &       19.64     \\
25      &       311528  &       203615  &       34.64     \\
50      &       313591  &       150919  &       51.87    \\
60      &       308210  &       127683  &       58.57     \\
70      &       295038  &       101134  &       65.72     \\
80      &       275313  &       72440   &       73.69     \\
90      &       238762  &       39710   &       83.37     \\
95      &       189995  &       21680   &       88.59      \\
99      &       78323   &       5167    &       93.40     \\  \hline
average & 557118 & 107342 & 63.28 \\ \hline
\end{tabular}
\caption{Results for random sequences of length $2^{20}$.}
\label{ta1}
\end{table}

\begin{table*}[t!]\centering\footnotesize
\begin{tabular}{|c|c|c|c|c|c||c|c|c|} \hline               
\multicolumn{6}{|c|}{Design parameters} & Bit-flipping & Presented & \multirow{2}{*}{$\frac{G_1-G_2}{G_1}*100\%$}  \\ \cline{1-6}
Name	& \# Gates & \# Scan cells & \# Faults & \# Scan chains & \# Top-off & \# Gates, $G_1$ & \# Gates, $G_2$ & \\ \hline
a	& 34113	&	2511	&	91834	&	128	&	207	&	5049	&	4565	&	9.59	\\
b	& 22104	&	1726	&	66390	&	128	&	214	&	4969	&	4515	&	9.14	\\
c	& 19621	&	1726	&	48920	&	128	&	40	&	973	&	472	&	51.52	\\
d	& 21984	&	1727	&	66554	&	128	&	211	&	5003	&	4470	&	10.65	\\
e	& 19275	&	1727	&	50076	&	128	&	66	&	1224	&	684	&	44.11	\\
f	& 39277	&	3022	&	89108	&	128	&	133	&	3820	&	3352	&	12.25	\\
g	& 31726	&	2690	&	83936	&	128	&	213	&	5125	&	4674	&	8.80	\\
h	& 29418	&	2690	&	67654	&	128	&	40	&	1191	&	577	&	51.55	\\ \hline
average & \multicolumn{5}{|c||}{}	& 3420	&	2880	&	24.7	\\ \hline
\end{tabular}
\caption{Comparison to the bit-flipping approach for maximum achievable stuck-at faults coverage.}
\label{ta2}
\end{table*}

\begin{table*}[th]\centering\footnotesize
\begin{tabular}{|c|c||c|c|c|c||c|c|c|c|} \hline 
\multicolumn{2}{|c|}{Design parameters}  & \multicolumn{4}{|c|}{Maximum achievable stuck-at fault coverage} & \multicolumn{4}{|c|}{Stuck-at fault coverage $> 98$\%} \\ \hline
\multirow{2}{*}{Name}	& \multirow{2}{*}{\# Gates, $G_1$} &  \multirow{2}{*}{\# Top-off} & \% Test &  Presented &   \multirow{2}{*}{$\frac{G_1-G_2}{G_1}*100\%$} &  \multirow{2}{*}{\# Top-off} & \% Test &  Presented &   \multirow{2}{*}{$\frac{G_1-G_3}{G_1}*100\%$} \\
	&               &       &  Coverage  & \# Gates, $G_2$ & &  & Coverage  & \# Gates, $G_3$ &   \\ \hline
a	& 34113 & 207 & 99.98 & 4565 & 13.29 & 102	& 98.21 & 2004 & 5.87 \\
b	& 22104	& 214	& 99.98	& 4515	& 20.28	& 63 & 98.30	& 1210	& 5.47	\\
c	& 19621 & 40 & 99.98	& 472 & 2.24	& 40 & 99.98 &	427	& 2.17	\\
d	& 21984	& 211	& 99.98	& 4470 & 20.18 & 67	& 99.13 & 1212 & 5.51 \\
e	& 19275 & 66	& 99.98	& 684 & 3.38	& 66	& 99.98 & 647 & 3.36		\\
f	& 39277 & 133 & 99.99 & 3352 & 8.45 & 100	& 99.46 & 2049 & 5.22 \\
g	& 31726 & 213 & 99.98 & 4674 & 14.63 & 92	& 99.89 & 1771 & 5.58 \\
h	& 29418	& 40 &	99.99	& 577 & 1.85	& 40	& 99.99 & 542 & 1.84	\\ \hline
\end{tabular}
\caption{Area overhead of the presented approach for different stuck-at fault coverages.}
\label{ta3}
\end{table*}

\subsection{Comparison to previous approaches for embedding deterministic test patterns}

In the second experiment, we compared the presented algorithms to the bit-flipping
approach for embedding deterministic test patterns which, in our 
opinion, is one of the most efficient ones~\cite{WuK96}. 
The results presented in this section were obtained using our implementation of 
the bit-flipping algorithm. 
 
\begin{figure}[t!]
\begin{center}
    \includegraphics*[clip=true,width=3.5in]{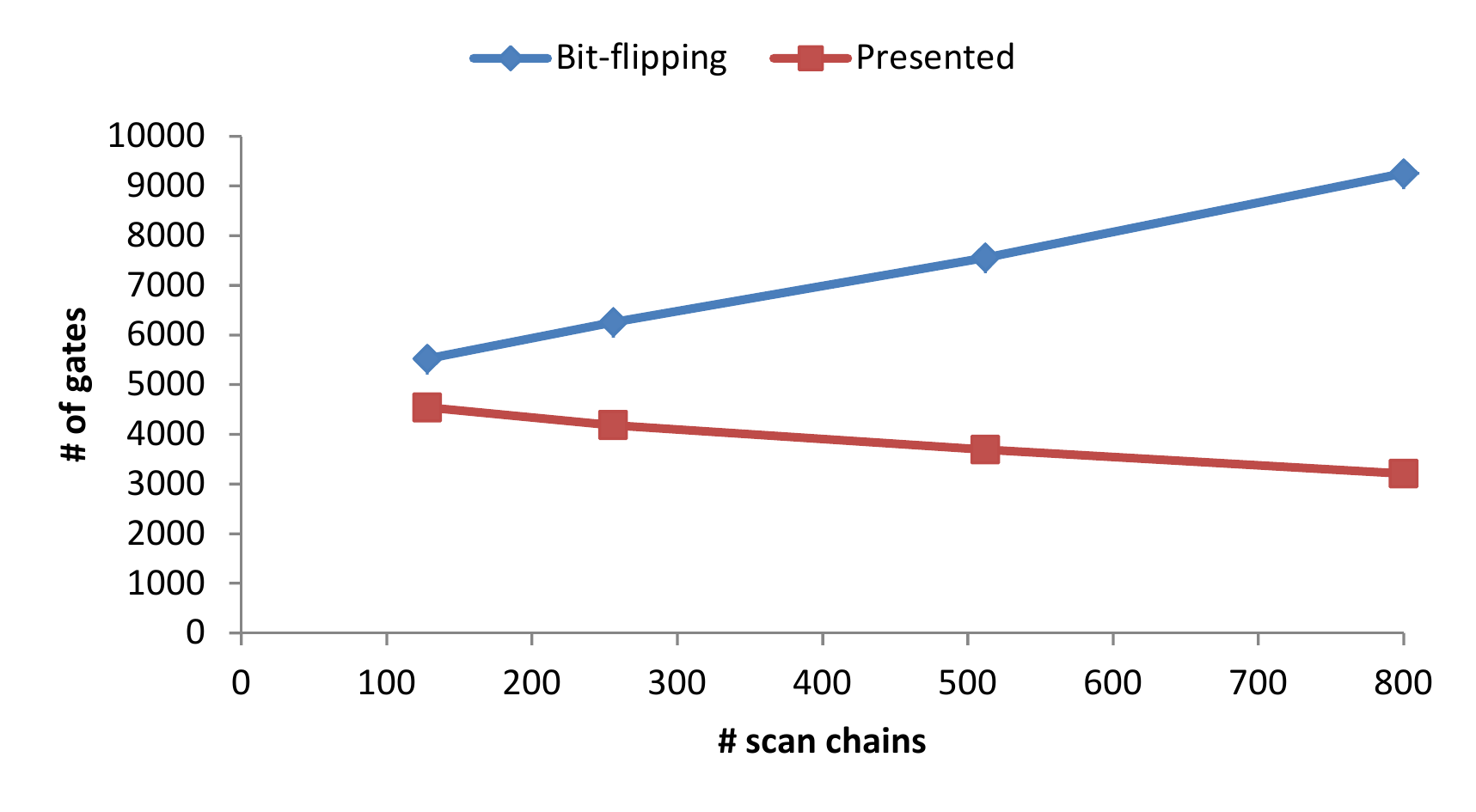}
\caption{Area overhead as a function of the number of chains.}\label{fs}
\end{center}
\end{figure}

We applied both algorithms to 8 real designs with the number of gates varying from
19K to 39K. The results are summarized in Table~\ref{ta2}. We first applied 9000 pseudo-random
patterns to all designs. Then, we computed the top-off patterns required
to reach maximum achievable stuck-at faults coverage using a commercial ATPG tool. 
We used bit-flipping and the presented algorithms to represent 
these top-off patterns. As we can see from  Table~\ref{ta2}, the presented
approach outperforms the bit-flipping approach by 24.7\% on average.
The difference in the number of gates required 
in both approaches can be up to 51.5\%.
What is even more important, the area overhead of the 
presented approach goes down as the number of scan chains grows. 
On the contrary, the area overhead of the bit-flipping approach goes up (see Figure~\ref{fs}).

However, in spite of the improvements,
the percentage of the overall chip area required to store deterministic test 
patterns can be prohibitively high for some designs (see column 6 of Table~\ref{ta3}).
It is known that the size of representation for a data is related to the entropy of data~\cite{Sh48}. 
Entropy puts a theoretical limit on the size of the minimal representation that can be achieved. 

\begin{figure}[t!]
\begin{center}
    \includegraphics*[clip=true,width=3.5in]{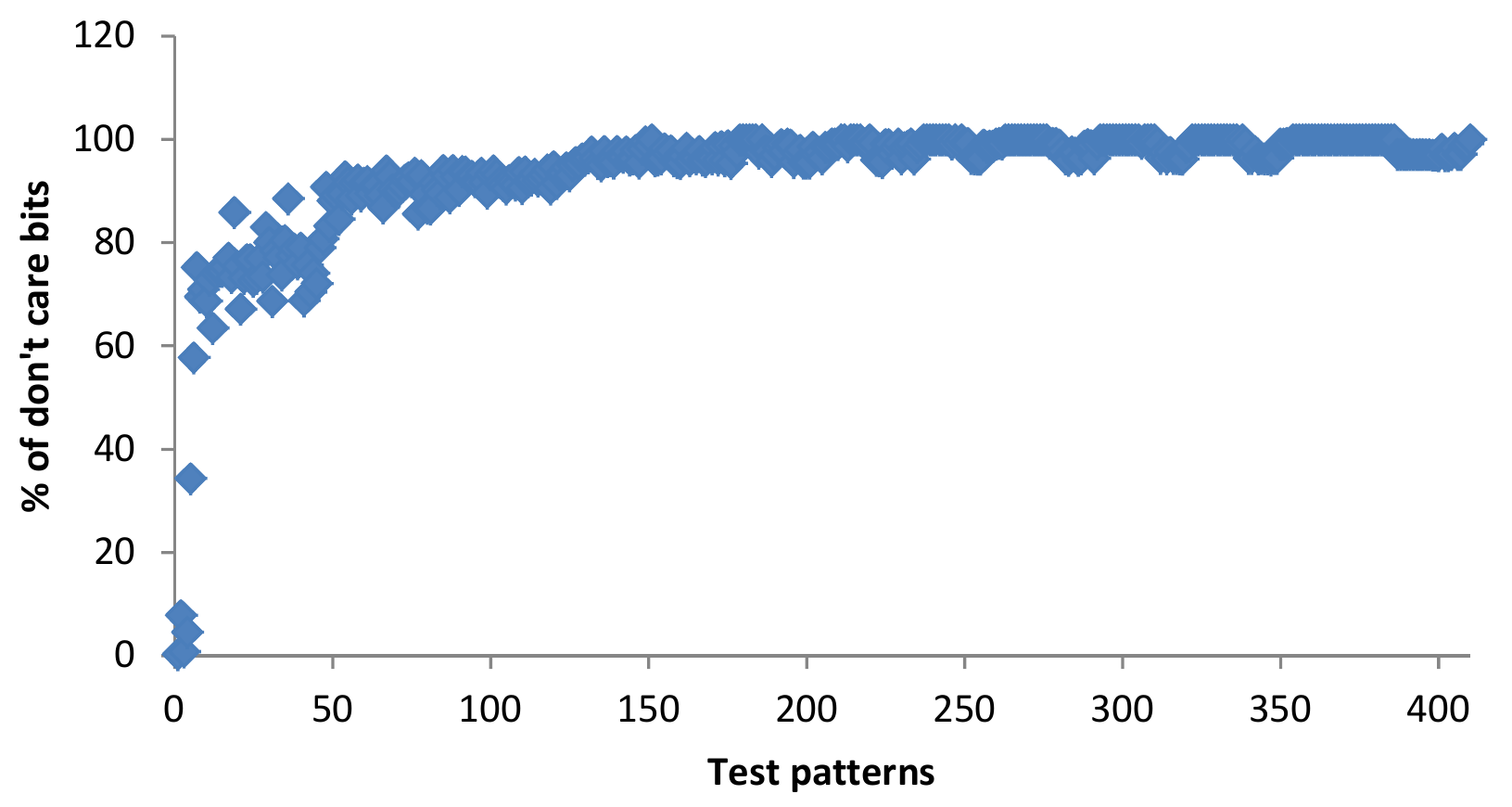}
\caption{Distribution of don't care bits in the test patterns of {\em dma} benchmark.}\label{dma}
\end{center}
\end{figure}

If a lower fault coverage is acceptable, then the area overhead can be 
reduced by exploiting the fact that don't care bits are normally unevenly
distributed among test patterns. As an example, consider the
diagram in Figure~\ref{dma}. Each point on this diagram shows the number of 
don't care bits in a test pattern of {\em dma} benchmark (in total 411 patterns of length 1720 bits each). 
These test patterns were generated by a commercial ATPG tool
with dynamic compaction turned on and random fill turned off.
They cover 100\% of detectable stuck-at faults.
The total percentage of specified bits is 6.45\%.
We can see that only the first few test patterns are highly specified.  
If we chop off the first 5\% of test patterns, the entropy of the remaining patterns
reduces twice. Therefore, they can be represented with 
a twice smaller representation than the one required for the whole set of test patterns.
By using the last 95\% of test patterns, we can achive 95.7\% test coverage for stuck-at faults.

In Table~\ref{ta3} we show that, by using a subset of the top-off patterns only,
we can reduce the area required for their representation more than 3 times in some cases,
while sacrificing the fault coverage by less than 2\%.

\section{Conclusion} \label{con}

We presented a new method for embedding deterministic test patterns on-chip
based on binary machines. The
presented algorithm for synthesis of binary machines is significantly 
more efficient than previous work, especially for test data with many
don't cares. 
Our experimental results on 8 real designs show that the proposed
approach outperforms the bit-flipping approach by 24.7\% on average.
We also show that it is possible to exploit uneven distribution of 
don't care bits in test patterns to reduce the area required for 
generating top-off patterns more than 3 times with less than 2\% 
decrease in fault coverage.

We believe that the presented algorithm for synthesis of binary machines
is quite close to an optimal. What can be improved in the proposed method 
is the strategy for selecting a subset of top-off patterns which
maximizes the fault coverage and minimizes the area overhead.
At present, we use a simple greedy algorithm which selects top-off patterns based on 
the number of don't care bits and the number of covered faults.
A more sophisticated approach is likely to bring better results.

Binary machines can potentially be used for storing
compressed test patterns for on-chip test compression techniques. 
This would eliminate the dependence of test compression 
on ATE memory. We are currently investigating the feasibility of
such an approach on large industrial designs.

\section{Acknowledgement} \label{ack}

This work was supported in part by a research grant No 2011-03336 from the Swedish Governmental Agency for Innovation Systems VINNOVA.

\bibliographystyle{ieeetr}
\bibliography{bib}

\end{document}